\documentclass[twocolumn,showpacs,preprintnumbers,superscriptaddress,amsmath,amssymb,pra]{revtex4}
\usepackage{epsfig}
\newcommand{\bm}[1]{ \mbox{\boldmath $#1$}  }

\begin{document}

\title{BCS-BEC Crossover in Bilayers of Cold Fermionic Polar Molecules}

\author{N.~T. Zinner}
\affiliation{Department of Physics, Harvard University, Cambridge MA, 02138}
\affiliation{Department of Physics and Astronomy, Aarhus University, Aarhus C, DK-8000}
\affiliation{The Niels Bohr Institute, University of Copenhagen, Copenhagen \O, DK-8000}
\author{B. Wunsch}
\affiliation{Department of Physics, Harvard University, Cambridge MA, 02138}
\author{D. Pekker}
\affiliation{Department of Physics, Harvard University, Cambridge MA, 02138}
\affiliation{Department of Physics, California Institute of Technology, Pasadena, California 91125, USA}
\author{D.-W. Wang}
\affiliation{Physics Department, National Tsing-Hua University, Hsinchu 300, Taiwan}
\affiliation{Physics Division, National Center for Theoretical Sciences, Hsinchu 300, Taiwan}
\affiliation{Frontier Research Center on Fundamental and Applied Sciences of Matters, National Tsing-Hua University, Hsinchu 300, Taiwan}
\date{\today}

\begin{abstract}
We investigate the quantum and thermal phase diagram of fermionic polar molecules loaded in a 
bilayer trapping potential with perpendicular dipole moment. We use both a BCS theory approach that 
is most realiable at weak-coupling and a strong-coupling approach that considers the two-body 
bound dimer states with one molecules in each layer as the relevant degree of freedom.
The system ground state is a Bose-Einstein condensate (BEC) of dimer bound states in the low density limit and a paired superfluid (BCS) state in the high density limit. 
At zero temperature, the intralayer repulsion is found to broaden the regime of BCS-BEC crossover, and can potentially induce system collapse through the softening of roton excitations. The BCS theory and the strongly-coupled dimer picture yield similar
predictions for the parameters of the crossover regime.
The BKT transition temperature of the dimer superfluid is also calculated. 
The crossover can be driven by many-body effects and is strongly affected by
the intralayer interaction which was ignored in previous studies.
\end{abstract}
\maketitle

\section{Introduction}
\begin{figure}[ht!]
  \epsfig{file=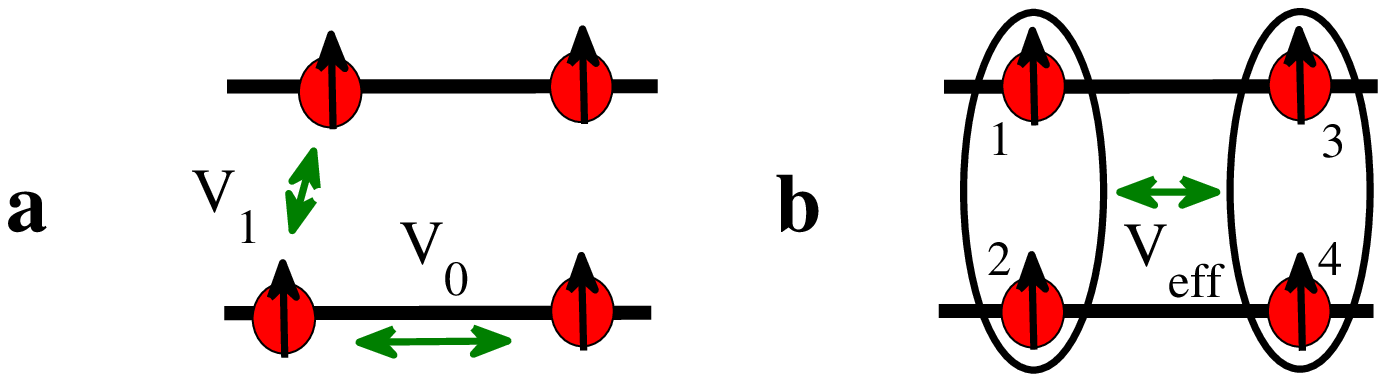,clip=true,scale=0.45,bb= 38   484   441   616}
  \epsfig{file=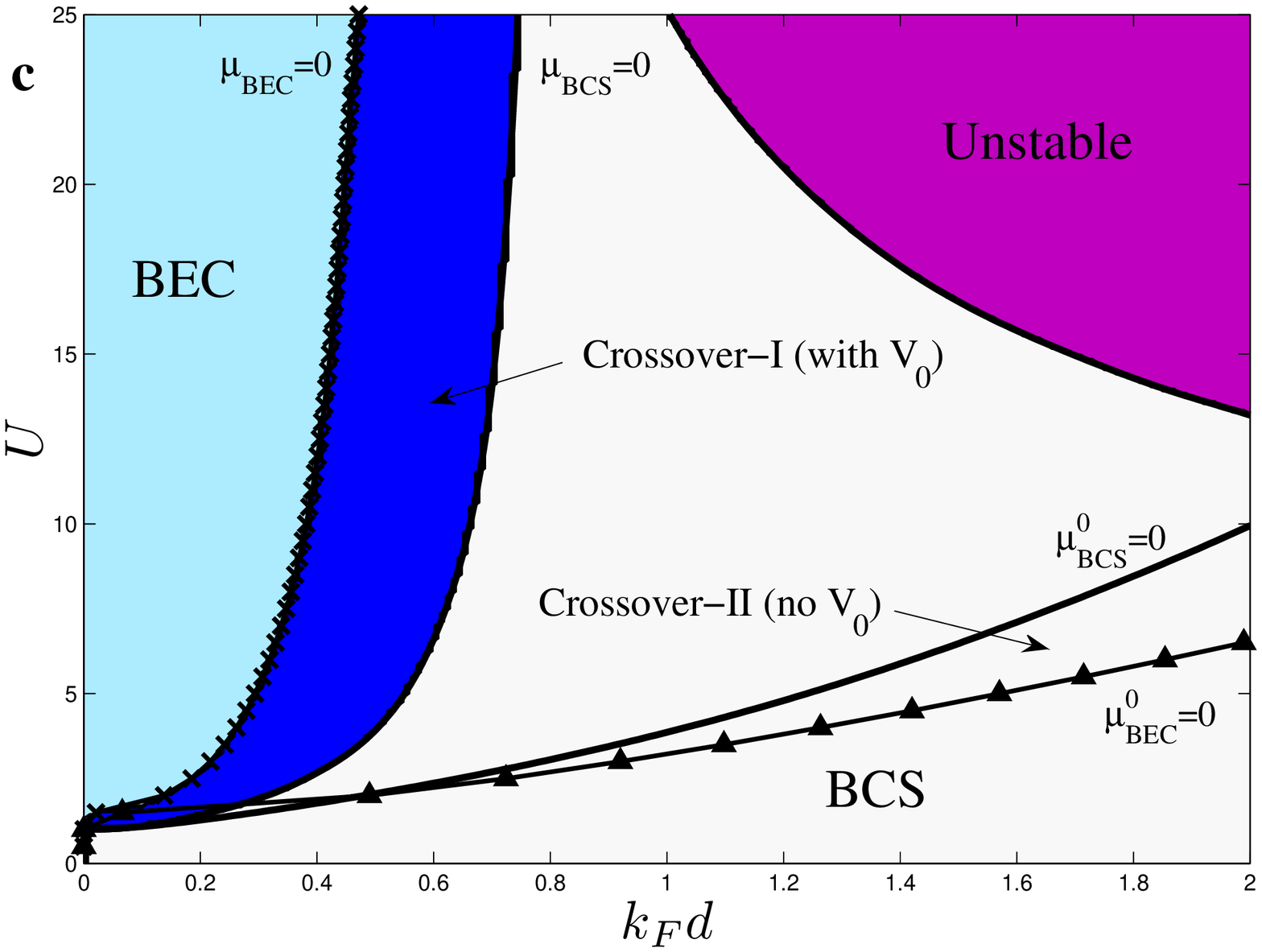,clip=true,scale=0.42}
  \caption{
(color online) 
Schematic of a few particles in the bilayer with
(a) inter- ($V_1$) and intralayer ($V_0$) interactions in the BCS 
limit and (b) the effective interaction ($V_{\rm eff}$) between 
the dimers in the BEC limit (see text).
(c) Quantum phase diagram for a bilayer with fermionic polar 
molecules in the interaction-density plane.
Interactions are characterized by $U=mD^2/\hbar^2d$ and by $k_Fd$, 
where $m$, $D$, $d$, and $k_F$ are respectively the molecular mass, 
dipole moment, interlayer distance, and Fermi momentum. The Crossover-I region 
includes the effect of intralayer interactions, while the Crossover-II region 
does not. In the upper right-hand corner we speculate that the system could 
potentially display a roton instability as discussed in the text.}
  \label{fig-zerophase}
\end{figure}

Recent progress on
trapping and cooling of polar molecules \cite{ospelkaus2008,ni2008,deiglmayr2008,lang2008,ni2010,ospelkaus2010} enable
studies of many-body systems with long-range anisotropic dipole-dipole forces, where
new exotic phases could exist \cite{baranov2008,lahaye2009}. 
The attractive part of the interaction can, however, lead to chemical reaction losses \cite{ospelkaus2010}.
A way to stabilize the system is to load molecules in a
one- or two-dimensional optical lattices, where interesting low-dimensional physics
has been predicted \cite{wang2006,wang2007,bruun2008,cooper2009,lutchyn2010,klawunn2010,mora2007,buchler2007,astra2007}.
A prominent example is polar fermions loaded in a bilayer system with dipoles oriented 
perpendicular to the layer plane \cite{miranda2010}. In the limit of high density or weak interaction, 
the system is very similar to conventional superconductors, and the ground state should be a BCS state (with interactions as in Fig.~\ref{fig-zerophase}(a)). In the dilute limit, it is known that the interlayer interaction always supports a bound state \cite{shih2009,armstrong2010,klawunn2010-2}, and the ground state should be a BEC of dimers (Fig.~\ref{fig-zerophase}(b)). As a result, the BCS-BEC crossover in this bilayer system could be richer than the usual atomic Fermi gas where crossover is driven by the two-body physics of a Feshbach resonance \cite{ketterle2008}. Here it is driven by many-body effects which depend not only on interaction strength but also on density. Furthermore, the intralayer repulsion can cause roton softening and/or Wigner crystallization in the high density limit \cite{mora2007,buchler2007,astra2007}. Therefore, a many-body theory including both intra- and interlayer interactions is not only of quantitatively interest but also qualitatively important.

In this paper, we study the quantum and thermal phase diagrams of fermionic polar molecules loaded in a bilayer system including both {\it intra}- and {\it inter}-layer interaction. The quantum phase diagram is shown in Fig.~\ref{fig-zerophase}(c).
BCS (BEC) ground states are found in the limit of weak (strong) interaction and large (small) density. In between, we have the crossover regime (Crossover-I), which can be determined by the chemical potential calculated from many-body theories in the different limits (see below).
When neglecting intralayer repulsion, the crossover region (Crossover-II) moves to lower interaction strength.
In addition, intralayer interactions could perhaps give rise to a roton instability at large density and strong interaction, although
further analysis beyond the scope of this work is needed. We also determine the BKT critical temperature \cite{bkt1,bkt2} in the strongly-coupled regime
including the effective interaction of the dimers. The maximum critical temperature obtained is one tenth of the Fermi energy, and should therefore be achievable in experimental setups in the near future.

We note that two other recent studies have considered a system similar to the one studied here. The paper by Pikovski {\it et al.} \cite{pikovski2010}
considers the BCS and BEC phases based on BCS theory at both zero and finite temperature. However, they do not consider the full effect of the intralayer interaction. When we neglect the intralayer interaction, our results are consistent with Ref.~\cite{pikovski2010}. A related study by 
Baranov {\it et al.} \cite{baranov2011} addresses the critical temperature for the superfluid phase in the weak-coupling limit, taking also 
particle-hole correlations into account. In contrast, here we consider the finite temperature phase diagram from the strong-coupling limit. Extrapolation of our results to the parameter regime of Ref.~\cite{baranov2011} would exceed the boundaries of our approximations and 
the current study should be viewed as complementary to Ref.~\cite{baranov2011}.

Our model and the assumptions we use are described Section~\ref{model}. In Section~\ref{zero}, we discuss the case of zero
temperature and the role of intralayer interactions on the BCS-BEC crossover. This is achieved by considering the physics
from both a weak- and a strong-coupling point of view. Both approaches are shown to yield consistent results. We 
proceed to discuss the finite temperature phase diagram in Section~\ref{finite}. Here we calculate the critical temperature
for the superfluid phase within mean-field theory and using the universal relation for the BKT transition temperature. 
Section~\ref{diss} contains a summary, a discussion of experimental parameters to realize the predicted
phases, and an outlook for future work.

\section{Model}\label{model}
The Hamiltonian for polar molecules in the bilayer system is given by
$H=\sum_{\bm k\sigma}\epsilon_{\bm k} c_{\bm k,\sigma}^{\dagger} c_{\bm k,\sigma}
+\frac{1}{2\Omega}\sum_{\bm q \bm k \bm k', \sigma \sigma' } V_{\sigma\sigma'}(\bm q)c_{\bm k+\bm q, \sigma}^{\dagger}
c_{\bm k'-\bm q, \sigma'}^{\dagger}c_{\bm k', \sigma'}c_{\bm k, \sigma}$,
where $\Omega$ is the area of the layer plane; $\epsilon_{\bm k}=\hbar^2\bm k^2/2m$, and $\sigma=\pm$ is the layer index. $V_{+-}=V_{-+}=V_1$ denotes the {\it inter}layer interaction and $V_{++}=V_{--}=V_0$ is the {\it intra}layer one. Here we neglect interlayer tunneling.
We assume occupation of only the ground state in the transverse direction so that the 
transverse degree of freedom is a simple gaussian that can be integrated out. This yields
$V_0(\bm q)=\frac{8\pi D^2}{3\sqrt{2\pi}W}\left(1-\tfrac{3}{2}F(|\bm q|)\right)$ and
$V_{1}(\bm q)\to -2\pi D^2|\bm q|e^{-|\bm q|d}$ as $W/d\to 0$. 
Here $\bm q$ is the in-plane momentum, $D$ is the dipole moment,
$W$ and $d$ the layer width and interlayer spacing, and $F(q)=\sqrt{\pi/2}Wq[1-\text{Erf}(Wq/\sqrt{2})]e^{q^2W^2/2}$ with
$\text{Erf}(x)$ the error function.
The interlayer interaction is exact in the strict 2D limit ($W\ll d$), but we find it is
accurate enough ($<10\%$) at $W=0.2d$, which is the value used throughout. 
In the weak interaction limit, the intralayer repulsion should renormalize the single particle dispersion as in Fermi liquid theory. Treating the layer index as a spin degree of freedom, the gap equation is
\begin{eqnarray}
\Delta_{\bm k}=-\frac{1}{\Omega}\sum_{\bm q} \frac{V_1(\bm k-\bm q)\Delta_{\bm q}}{2E_{\bm q}}\text{tanh}\left(\frac{E_{\bm q}}{2k_BT}\right),
\label{gap}
\end{eqnarray}
where $E_{\bm q}=\sqrt{\xi_{\bm q}^{2}+|\Delta_{\bm q}|^{2}}$ is the quasi-particle dispersion, and
$\xi_{\bm q}\equiv\epsilon_{\bm q}+\Sigma(\bm q)-\mu_{BCS}$. The self-energy, $\Sigma({\bm q})$, within in the Hartree-Fock approximation is \cite{HFB}
$\Sigma(\bm k)=\frac{1}{2\Omega}\sum_{\bm q}(V_0(0)-V_0(\bm k-\bm q))
[1-\frac{\xi_{\bm q}}{E_{\bm q}}\text{tanh}(E_{\bm q}/2k_BT)]$.
To access the crossover regime, the chemical potential, $\mu_{BCS}$, must be determined self-consistently via the density equation
\begin{align}
n=\frac{1}{2\Omega}\sum_{\bm q}\left[1-\frac{\xi_{\bm q}}{E_{\bm q}}
\text{tanh}\left(\frac{E_{\bm q}}{2k_BT}\right)\right].
\label{density}
\end{align}
We use the first Born approximation (FBA) with a realistic finite
layer width for both intra- and interlayer interactions in Eq.~\eqref{gap}. 
The exponentially decreasing shape of $V_1(\bm q)$ means that using 
the renormalized gap equations or the FBA yields similar results
as noted already in Ref.~\cite{pikovski2010}.
The FBA is generally poor at strong coupling. However,
since the crossover takes place at low density, we treat
the intralayer in a weak coupling sense in the current work. Inclusion 
of higher-order terms will be considered in future work.

Since the interlayer interaction, $V_1$, is attrative at short distance, the dominant gap is $s$-wave. Therefore, for simplicity, we neglect higher partial wave components and calculate $\Delta_{\bm k}=\Delta_{|\bm k|}$ from Eq.~\eqref{gap} by iteration. In
Fig.~\ref{fig-spec}(a), we show $\Delta_k/\Delta_0$ for different
values of the dimensionless coupling, $U=mD^2/\hbar^2 d$ with $k_Fd=\sqrt{4\pi nd^2}=0.4$. When $U$ is small, the maximum of $\Delta(k)$ is at $k/k_F=2.5$, which is where $V_1(k)$ is most attractive. When $U$ is larger, the intralayer interaction renormalizes the single particle dispersion; the structure of $\Delta_k$ disappears and the maximum is at $k=0$. In Fig.~\ref{fig-spec}(b) we show $\Delta_0(T)/E_F$ as function of $T/T_F$, where $E_F$ and $T_F$ are Fermi energy and temperature.

\section{Zero Temperature Results: Intralayer Interaction Effects on the BCS-BEC Crossover}\label{zero}
\begin{figure}[t!]
  \epsfig{file=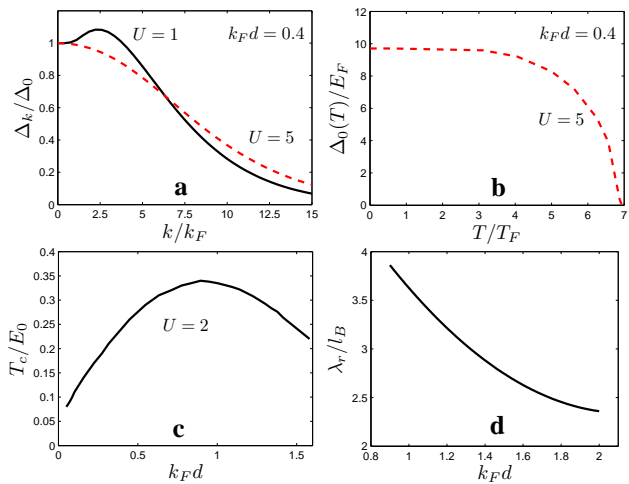,clip=true,scale=0.42}
  \caption{(color online) (a) Normalized gap, $\Delta_k/\Delta_0$, as a function of $k/k_F$ at $k_Fd=0.4$ and $U=1$ (solid) and $U=5$ (dashed). (b) Temperature dependence of $\Delta_0(T)/E_F$ as a function of $T/T_F$ for $k_Fd=0.4$ ($E_F=0.16 E_0$).
(c) BCS critical temperature as a function of $k_Fd$ at $U=2$ in units of $E_0=\hbar^2/2md^2$. (d) Ratio of the roton wavelength, $\lambda_r$, at the instability to the dimer size, $l_B$, as function of $k_Fd$.}
  \label{fig-spec}
\end{figure}

When the density is reduced, the ground state is expected to be a dimer BEC. Within BCS theory, this regime can be defined by having a negative chemical potential ($\mu_{BCS}<0$). From Eq.~\eqref{density}, we can determine the $\mu_{BCS}=0$ boundary as shown in Fig.~\ref{fig-zerophase}(c).
The two solid lines that bound the two crossover regions are calculated by including ($\mu_{BCS}$) and neglecting ($\mu_{BCS}^{0}$) the intralayer repulsion respectively. As expected, the repulsion
strongly suppresses the $\mu_{BCS}<0$ region and for densities larger than a critical value of $k_Fd\sim 0.77$,
$\mu_{BCS}$ is positive for any $U$. We always find a $\mu_{BCS}^{0}<0$ region when the intralayer interaction is neglected. This shows that the intralayer interaction brings not only quantitative contributions, but also qualitative and important changes of the quantum phase diagram. The effect of intralayer repulsion on the crossover physics can also be seen in the transition temperature, $T_c$, as shown in Fig.~\ref{fig-spec}(c). It has a maximum at $k_Fd\sim 1$, similar to crossover in a quasi-2D superconductor \cite{li2001}. In contrast to BCS results for Fermi gases with short-range interaction, the decrease of $T_c$ at high density is caused by the long-range intralayer repulsion. 

In order to investigate the many-body physics of the BEC limit (i.e. the strong interaction or low density limit) in more detail, we go beyond BCS theory, which is based on Fermi liquid theory at high density.
Starting from the extremely dilute limit, the bound dimer is the main constituent with binding energy $E_B$. For small density, we neglect the intralayer term but include the Fermi pressure in the chemical potential; $\mu_{BEC}^0=E_F-E_B/2$, where $E_B$ is obtained numerically \cite{shih2009,armstrong2010,klawunn2010-2}. We define the crossover regime (Crossover-II) as the region bounded by $\mu_{BCS}^{0}=0$
and $\mu_{BEC}^0=0$, the latter given by the line with triangles in Fig.~\ref{fig-zerophase}(c). The Crossover-II region is quite narrow which indicates that the BCS theory and the strong-coupling result are very similar. The expression for $\mu_{BCS}^{0}$ given above holds exactly for zero-range interactions \cite{randeria1990}, and indicates that the crossover happens when the dimer size becomes comparable to the interparticle distance. While the interlayer dipole interaction behaves similar to a zero-range interaction, the intralayer dimer-dimer interaction can have significant effects on $\mu_{BEC}$ as we now demonstrate by deriving an effective interaction between dimers. We note that a recent Monte Carlo study of the BCS-BEC crossover in two dimensions also finds that in the BEC limit, the dimer-dimer and atom-dimer interactions are important corrections that are not taken into account in the usual BCS theory without self-energy corrections \cite{bertaina2011}.

The coordinates of the four molecules are denoted by $\bm r_1,\cdots,\bm r_4$ as shown in Fig. \ref{fig-zerophase}(b), where $(\bm r_1,\bm r_2)$ are for the left dimer and $(\bm r_3, \bm r_4)$ for the right dimer.
We are interested in deep bound state, where the dimer size is smaller than the inter-dimer distance, i.e. $|\bm \rho|\gg |\bm r|,|\bm r'|$, where
$\bm \rho=(\bm r_1+\bm r_2)/2-(\bm r_3+\bm r_4)/2$ is inter-dimer distance, and $\bm r=\bm r_1-\bm r_2$ and $\bm r'=\bm r_3-\bm r_4$ are relative coordinates in each dimer. Straighforward algebra and integration over the dimer bound state wavefunction, $\phi(\bm r)$, yields the effective dimer-dimer interaction
$V_{\text{eff}}(\bm \rho)=\int\,d\bm r\,d\bm r'\, |\phi(\bm r)|^2|\phi(\bm r')|^2
\sum_{s=\pm 1,\alpha=0,1}V_\alpha(\bm \rho+s\bm r_\alpha)$,
where $\bm r_{0,1}\equiv (\bm r\mp\bm r')/2$. 
In the strong interaction and dilute density limit, we can approximate $\phi(\bm r)$ by a Gaussian profile; 
$\phi(\bm r)=(l_B\sqrt{\pi})^{-1}\exp\left(-|\bm r|^2/2l_B^2\right)$, where $l_B=\sqrt{2\hbar^2/mE_B}$ is the radius of dimer bound state. As a result, $V_{\rm eff}$ has the following simple Fourier transform:
\begin{align}
V_{\text{eff}}(\bm k)=[2V_1(\bm k)+2V_0(\bm k)]\exp(-|\bm k|^2l_B^2/8).
\end{align}
The Gaussian approximation for $\phi(\bm r)$ fails for $U\lesssim 2$, however
we have checked that our results are qualitatively unchanged if the exact solution is used \cite{zinner2011}.
Notice that $V_{\text{eff}}$ takes the strong interlayer interaction into account through the dimer wavefunction. 
The FBA is used for the intralayer term, which is reasonable since the crossover happens 
at low density. For the roton instability, the FBA was used in quasi-2D studies of dipolar bosons \cite{santos2003,fischer2006}
and we expect it to remain a fair approximation at larger densities as well. We note
that dipolar interactions are different from short- or zero-range interactions.
Zero-range interactions in 2D always allow a two-body dimer bound state \cite{randeria1990}, and 
a state of four bosons will also be bound \cite{brodsky2006}. This result is different from 3D 
where the interaction must be sufficiently attractive to produce bound states. In the current 
setup, the dimer-dimer system is unbound \cite{artem2011c}.

The effective interaction has the property that $V_{\text{eff}}(0)$ is non-zero due to the intralayer term.
However, the molecules are fermionic and the Fock exchange contribution could cancel this term as is 
the case with true short-range interactions in interacting single-component Fermi gases. However, 
$V_{\text{eff}}(\bm q)$ also has large contributions from $\bm qd>0$, and this should have an
influence on the phase diagram. This is supported by recent studies of the density-wave instability
where the effect of exchange is found to be very large, shifting the instability into the strong-coupling
regime \cite{zinnerbruun2011,babadi2011,parish2011,sieberer2011,kusk2011}. The importance of 
exchange effects has also been discussed in relation to ferroeletricity with polar 
molecules \cite{lin2010}.
Here we estimate the effects in 
the bosonic dimer limit by including the intralayer potential through $V_{\text{eff}}(0)$.

Using $V_{\text{eff}}$, the chemical potential in the BEC limit can be estimated as $\mu_{BEC}=nV_{\text{eff}}(0)/2-E_B/2$
(we neglect the Fermi pressure which is much smaller than the interaction energy for strong interactions).
In Fig.~\ref{fig-zerophase}(c), the solid line with crosses is given by $\mu_{BEC}=0$ and is the upper bound of the crossover regime including intralayer interaction (Crossover-I). Notice in particular that the BEC region shrinks to lower density when including the intralayer interaction, and there is no dimer condensate for $k_Fd\gtrsim 0.42$.
The ratio of interparticle distance to dimer size is five or more
along the $\mu_{BCS}=0$ and $\mu_{BEC}=0$ lines. Without intralayer interaction the ratio is around 2.5, again demonstrating 
how the crossover physics is strongly modified by the intralayer term.
We have estimate the chemical potential from both BCS and BEC limits, giving bounds of the crossover regime with and without the intralayer interaction. Our results indicate that in a realistic experiment, the intralayer repulsion can significantly affect the regime where a dimer condensate is observable.  

From the effective interaction between dimers, $V_{\text{eff}}$, we can also calculate the dispersion, 
$\hbar\omega_d({\bm k})=\sqrt{\epsilon_{\bm k}/2(\epsilon_{\bm k}/2+2nV_{\text{eff}}(\bm k))}$, of the collective Bogoliubov mode of a dimer superfluid in analogy to the case of dipolar bosons \cite{fischer2006}.
With increasing $U$ we find roton softening around $kd=2\pi$, which corresponds to a wavelength of $\lambda_r=2\pi/k\sim d$. 
This softening leads to system collapse in the high density and strong interaction regime, as shown in Fig.~\ref{fig-zerophase}(c). 
To investigate the nature of the instability, we plot $\lambda_r/l_B$ in Fig.~\ref{fig-spec}(d) and observe that $\lambda_r$ is more than a factor of two larger than $l_B$ for all densities. This implies that, at least in the low-density limit, it is a many-body effect.

The roton analysis assumes a well-defined dimer picture. We find the instability in the upper right-hand corner of Fig.~\ref{fig-zerophase}c, i.e. 
at higher densities and large $U$. While the dimer picture should prevail for large $U$, the higher densities imply that some
fermionic nature could perhaps arise.
Similar effects should arise in a fermionic picture but it is not easily calculated since fluctuations beyond BCS theory are needed. 
So at this point the findings for the roton instability remains speculative. This is an interesting topic for future work. 
We note that density waves in a single layer with 
fermionic polar molecules have been predicted recently \cite{sun2010,yamaguchi2010}. A rough estimate indicates that 
the roton instability lies inside the region where a density wave in single layers is predicted. However, Refs.~\cite{sun2010} and
\cite{yamaguchi2010} neglect the Fock contributions which are expected to be large \cite{zinnerbruun2011,babadi2011,parish2011,sieberer2011,kusk2011}. The exact 
region of the density wave instability is therefore not yet known. In any case, we expect the system to be unstable in the upper right part of the zero
temperature phase diagram of Fig.~\ref{fig-zerophase}.

\section{Finite Temperature Results: Intralayer Interaction Effects on the Critical Temperature}\label{finite}
\begin{figure}[t!]
  \epsfig{file=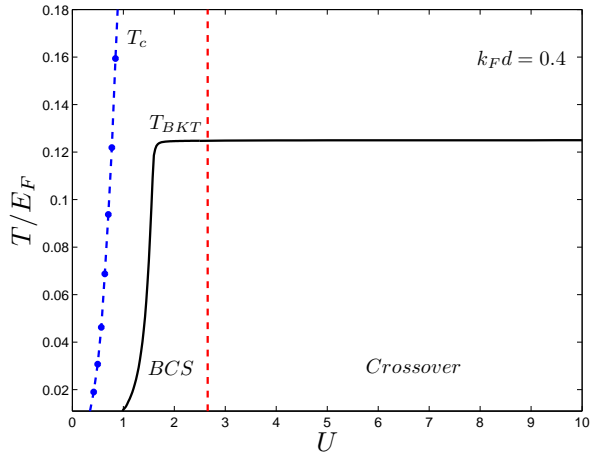,clip=true,scale=0.43}
  \caption{(color online) Transition temperature ($T_{BKT}$, solid black line) as a function of interaction strength for $k_Fd=0.4$. BCS result ($T_c$, dashed blue line) is shown for comparison. The dashed vertical line marks the position of $\mu_{BCS}=0$ (see Fig.~\ref{fig-zerophase}(c)).}
  \label{fig-temp}
\end{figure}

We now investigate the finite temperature phase diagram. In the BCS limit, we can use Eq.~\eqref{gap} to obtain the transition temperature, $T_c$, as shown by the dashed line in Fig.~\ref{fig-temp}. We note that the intralayer interaction has very little influence on $T_c$ at $k_Fd=0.4$.
It is known that in the weak interaction limit, $T_c$ is very close to the true transition temperature in a full BKT theory \cite{miyake1983}. 
However, this calculation fails in the strong interaction or dilute limit. At strong coupling, we obtain the BKT transition temperature from the univeral relation $k_B T_{BKT}= \hbar^2 \pi n_s(T_{BKT})/2m$, where $n_s(T)$ is the superfluid density at temperature $T$ \cite{bkt1,bkt2}. According to Landau's two-fluid model, we have $n_s(T)=n-n_n(T)$, with normal fluid density given by
$n_n(T)=\frac{\hbar^2}{16m k_B T}\int \frac{d^2q}{(2\pi)^2}\left[\frac{q}{\text{Sinh}(\omega_d(\bm q)/2k_B T)}\right]^2$ \cite{nelson1997}.
In Fig.~\ref{fig-temp} we show $T_{BKT}$ of a dimer superfluid as a function of $U$ for $k_Fd=0.4$ (solid black line). 
The dimer result differs from the BCS theory prediction in the weak coupling limit, while it becomes saturated at $T_{BKT}=0.125T_F$
for large $U$. In Ref. \cite{pikovski2010}, $T_c$ and $T_{BKT}$ was calculated from the BCS superfluid state. For small $U$ we find the same result for $T_c$. For dipolar bosons, Ref.~\cite{filinov2010} finds $T_{BKT}\sim 0.1T_F$ for all values of $U$ shown in Fig.~\ref{fig-temp}. 
In order to compare to that study we need to assume point dimers with twice the dipole moment, which is of course not the case for small $U$ where $T_{BKT}$ is reduced as the dimer size grows. Our strong-coupling dimer approach to the finite temperature physics can be considered complementary to both Ref. \cite{pikovski2010} and Ref. \cite{filinov2010}. The good agreement at intermediate and large $U$ with 
the other approaches provides support that we capture the essential physics using the effective interaction between dimers.

\section{Discussion}\label{diss}
We have studied fermionic polar molecules in a bilayer with perpendicularly polarized dipole moments. As the density and dipole strength varies, an analogue (but with different physics) of the celebrated BCS-BEC crossover is predicted. We find that the intralayer repulsion (and dimer-dimer interaction) is crucial quantitatively and also to some extend qualitatively: It shrinks the region of dimer condensation so that no dimer state is expected when the density is larger than a critical value and it also causes a roton instability in the high density and strong interaction regime. Our work is thus important for the study of BCS-BEC crossover physics in 2D systems, and our results should be observable within the parameter regime of near future experiments.

It is worth stressing that our study considers the physics of the bilayer with fermionic polar molecules from different points of view.
The BCS theory is usally more reliable in the weak-coupling regime, but can be extended into the crossover regime by solving the 
gap and number equations self-consistently. As the bilayer setup will always have a two-body bound state with one molecule in each
layer, it is reasonable to consider these dimers as the relevant degree of freedom in the strongly-coupled regime. The results presented
above indicate that both approaches yield consistent results, with or without including the repulsive intralayer interaction.

The crystalline phases \cite{mora2007,buchler2007,astra2007} are also ground state candidates due to the intralayer repulsion. For $W/d=0.2$ used in our calculations, these phases could appear below the roton instability region in Fig~\ref{fig-zerophase}. 
The finite extend of the dimers is, however, still a concern and further work is needed to determine crystal phases in the bilayer setup. 

In order to detect the phases a number of techniques could be applied. Dimerization in the layers should be detectable by Bragg scattering \cite{potter2010} or {\it in situ} non-demolition detection \cite{wunsch2011,wunsch2011b}, whereas RF spectroscopy can probe the gap. 
To probe the finite temperature physics one can detect the associated vortices by matter wave heteredyning \cite{kruger2006}. 
To estimate parameters for relevant systems, we take $d=0.5\mu$m which yields $U=1.22$ for KRb molecules at $D=0.566$ Debye and
$U=4.15$ for LiCs at $D=1.0$ Debye (the permanent dipole moment of this molecule in the ground-state is about $D=5.4$ Debye \cite{deigl2010}). For densities
$n=10^{5}-10^{8}$ cm$^{-2}$ we have $k_Fd=0.06-1.8$. The critical density to reach $\mu_{BCS}<0$ is about
$n\sim 0.2\cdot 10^{8}$ cm$^{-2}$, whereas the dimer BEC requires $n\lesssim 0.6\cdot 10^7$ cm$^{-2}$. Unfortunately, as $T_F=E_0 (k_Fd)^2/k_B$ this means that extremely low temperatures ($\lesssim 1$nK) are required to reach $T_{BKT}$.

Interesting directions for future work includes more than two layers or tilting of the dipoles with respect to the 
plane. With dipoles that are no longer perpendicular one can still show that a two-body bound dimer will be present
for any value of the dipole moment \cite{artem2011a}, although the dimer binding energy is reduced \cite{artem2011b}.
In the single layer case, $p$-wave superfluidity can occur for an extended range of tilting angles \cite{bruun2008}
and we expect similar effects for a bilayer. This setup can be explored with the same methods used here. 
In a setting with multiple layers more phases should be expected such as coherent density waves \cite{zinnerbruun2011}, 
pairing \cite{potter2010}, and bound states with more than two molecules \cite{jeremy2011,artem2011c}.

\acknowledgments
We thank E. Demler, L. Pollet, A.~S. Jensen, G.~M. Bruun, M.~M. Parish for fruitful
discussions and C.-H. Lin and I.-W. Tsai for early pioneering work. DWW appreciate the
hospitality of the JQI during the initial investigation of this project, and acknowledges
research and travel support from NCTS (Taiwan).
Support by the Carlsberg Foundation and by the German Science Foundation under grant WU 609/1-1 is gratefully acknowledged.


\begin{thebibliography}{99}

\bibitem{ospelkaus2008} S. Ospelkaus {\it et al.}, Nature Phys. {\bf 4}, 622 (2008).
\bibitem{ni2008} K.~K. Ni {\it et al.}, Science {\bf 322}, 231 (2008).
\bibitem{deiglmayr2008} J. Deiglmayr {\it et al.}, Phys. Rev. Lett. {\bf 101}, 133004 (2008).
\bibitem{lang2008} F. Lang, K. Winkler, C. Strauss, R. Grimm, and J.~H. Hecker Denschlag, Phys. Rev. Lett. {\bf 101}, 133005 (2008).
\bibitem{ni2010} K.~K. Ni {\it et al.}, Nature {\bf 464}, 1324 (2010).


\bibitem{ospelkaus2010} S. Ospelkaus {\it et al.}, Science {\bf 327}, 853 (2010).


\bibitem{baranov2008} M.~A. Baranov, Phys. Rep. {\bf 464}, 71 (2008).
\bibitem{lahaye2009} T. Lahaye, C. Menotti, L. Santos, M. Lewenstein, and T. Pfau, Rep. Prog. Phys. {\bf 72}, 126401 (2009).

\bibitem{wang2006} D.-W. Wang, M.~D. Lukin, and E. Demler, Phys. Rev. Lett. {\bf 97}, 180413 (2006).
\bibitem{wang2007} D.-W. Wang, Phys. Rev. Lett. {\bf 98}, 060403 (2007).
\bibitem{bruun2008} G.~M. Bruun and E. Taylor, Phys. Rev. Lett. {\bf 101}, 245301 (2008).
\bibitem{cooper2009} N.~R. Cooper and G.~V. Shlyapnikov, Phys. Rev. Lett. {\bf 103}, 155302 (2009).
\bibitem{lutchyn2010} R.~M. Lutchyn, E. Rossi, S. Das Sarma, Phys. Rev. A {\bf 82}, 061604(R) (2010).
\bibitem{klawunn2010} M. Klawunn, J. Duhme, and L. Santos, Phys. Rev. A {\bf 81}, 013604 (2010).

\bibitem{mora2007} C. Mora, O. Parcollet, and X. Waintal, Phys. Rev. B {\bf 76}, 064511 (2007).
\bibitem{buchler2007} H.-P. B{\"u}chler {\it et al.}, Phys. Rev. Lett. {\bf 98}, 060404 (2007).
\bibitem{astra2007} G.~E. Astrakharchik, J. Boronat, I.~L. Kurbakov, and Yu.~E. Lozovik, Phys. Rev. Lett. {\bf 98}, 060405 (2007).

\bibitem{miranda2010} M.~G.~H. de Miranda {\it et al.}, Nature Phys. {\bf 7}, 502 (2011).

\bibitem{shih2009} S.-M. Shih and D.-W. Wang, Phys. Rev. A {\bf 79}, 065603 (2009).
\bibitem{armstrong2010} J.~R. Armstrong, N.~T. Zinner, D.~V. Fedorov, and A.~S. Jensen, EPL {\bf 91}, 16001 (2010).
\bibitem{klawunn2010-2} M. Klawunn, A. Pikovski, and L. Santos, Phys. Rev. A {\bf 82}, 044701 (2010).

\bibitem{ketterle2008} W. Ketterle and M.~W. Zwierlein, in Ultracold Fermi Gases, Proceedings of the International School of Physics Enrico Fermi, Course CLXIV, Varenna, June 2006, edited by M. Inguscio, W. Ketterle, and C. Salomon (IOS Press, Amsterdam) 2008.


\bibitem{bkt1} V.~L. Berezinskii, Sov. Phys. JETP {\bf 34}, 610 (1972). 
\bibitem{bkt2} J.~M. Kosterlitz and D.~J. Thouless, J. Phys. C {\bf 6}, 1181 (1973).

\bibitem{pikovski2010} A. Pikovski, M. Klawunn, G.~V. Shlyapnikov, and L. Santos, Phys. Rev. Lett. {\bf 105}, 215302 (2010).

\bibitem{baranov2011} M.~A. Baranov, A. Micheli, S. Ronen, and P. Zoller, Phys. Rev. A {\bf 83}, 043602 (2011).

\bibitem{HFB} C. Zhao {\it et al.}, Phys. Rev. A {\bf 81}, 063642 (2010).

\bibitem{li2001} Z. Li and K. Yamada, JPSJ {\bf 70}, 797 (2001).

\bibitem{randeria1990} M. Randeria, J.-M. Duan, and L.-Y. Shieh, Phys. Rev. B {\bf 41}, 327 (1990).

\bibitem{bertaina2011} G. Bertaina and S. Giorgini, Phys. Rev. Lett. {\bf 106}, 110403 (2011).

\bibitem{zinner2011} N.~T. Zinner, J.~R. Armstrong, A.~G. Volosniev, D.~V. Fedorov, and A.~S. Jensen, arXiv:1105.6264v1.

\bibitem{santos2003} L Santos, G.~V. Shlyapnikov, and M. Lewenstein, Phys. Rev. Lett. {\bf 90}, 250403 (2003).

\bibitem{fischer2006} U.~R. Fischer, Phys. Rev. A {\bf 73}, 031602(R) (2006).

\bibitem{brodsky2006} I. V. Brodsky, M. Yu. Kagan, A. V. Klaptsov, R. Combescot, and X. Leyronas, Phys. Rev. A {\bf 73}, 032724 (2006).

\bibitem{artem2011c} A.~G. Volosniev, D.~V. Fedorov, A.~S. Jensen, and N.~T. Zinner, arXiv:1109.4602v1.

\bibitem{zinnerbruun2011} N.~T. Zinner and G.~M. Bruun, Eur. Phys. J. D {\bf 65}, 133 (2011).
\bibitem{babadi2011} M. Babadi and E. Demler, Phys. Rev. B {\bf 84}, 235124 (2011).
\bibitem{sieberer2011} L.~M. Sieberer and M.~A. Baranov, arXiv:1110.3679v1
\bibitem{parish2011} M.~M. Parish and F.~M. Marchetti, arXiv:1109.2464v1
\bibitem{kusk2011} J.~K. Block, N.~T. Zinner, and G.~M. Bruun, in preparation.
\bibitem{lin2010} C.-H. Lin, Y.-T. Hsu, H. Lee, and D.-W. Wang, Phys. Rev. A {\bf 81}, 031601(R) (2010).

\bibitem{sun2010} K. Sun, C. Wu, and S. Das Sarma, Phys. Rev. B {\bf 82}, 075105 (2010).
\bibitem{yamaguchi2010} Y. Yamaguchi, T. Sogo, T. Ito, and T. Miyakawa, Phys. Rev. A {\bf 82}, 013643 (2010).

\bibitem{miyake1983} K. Miyake, Prog. Theor. Phys. {\bf 69}, 1794 (1983).

\bibitem{nelson1997} U.~C. Tauber and D.~R. Nelson, Phys. Rep. {\bf 289}, 157 (1997).

\bibitem{filinov2010} A. Filinov, N.~V. Prokof'ev, and M. Bonitz, Phys. Rev. Lett. {\bf 105}, 070401 (2010).

\bibitem{potter2010} A.~C. Potter, E. Berg, D.-W. Wang, B.~I. Halperin, and E. Demler, Phys. Rev. Lett. {\bf 105}, 220406 (2010).

\bibitem{wunsch2011} B. Wunsch, N.~T. Zinner, I.~B. Mekhov, S.-J. Huang, D.-W. Wang, and E. Demler, Phys. Rev. Lett. {\bf 107}, 073201 (2011).
\bibitem{wunsch2011b} N.~T. Zinner, B. Wunsch, I.~B. Mekhov, S.-J. Huang, D.-W, Wang, and E. Demler, Phys. Rev. A {\bf 84}, 063606 (2011).
\bibitem{kruger2006} P. Kr{\"u}ger {\it et al.}, Nature {\bf 441}, 1118 (2006).

\bibitem{deigl2010} J. Deiglmayr, A. Grochola, M. Repp, O. Dulieu, R. Wester, and M. Weidem{\"u}ller, Phys. Rev. A {\bf 82}, 032503 (2010).

\bibitem{artem2011a} A.~G. Volosniev, D.~V. Fedorov, A.~S. Jensen, and N.~T. Zinner, Phys. Rev. Lett. {\bf 106}, 250401 (2011).
\bibitem{artem2011b} A.~G. Volosniev, N.~T. Zinner, D.~V. Fedorov, A.~S. Jensen, and B. Wunsch, J. Phys. B {\bf 44} (2011) 125301.
\bibitem{jeremy2011} J.~R. Armstrong, N.~T. Zinner, D.~V. Fedorov, and A.~S. Jensen, arXiv:1106.2102v1.

\end{thebibliography}
\end{document}